# Biocarbon reinforced polypropylene composite: An investigation of mechanical and filler behavior through advanced dynamic atomic force microscopy and X-ray micro CT

*J. George*[1*], *D. Bhattacharyya*[1,2]

[1]Centre for Advanced Composite Materials, Faculty of Engineering, The University of Auckland, 314 Khyber Pass Road, Newmarket, 1142 Auckland, New Zealand
[2]Plastics Centre of Excellence, Faculty of Engineering, The University of Auckland, 314 Khyber Pass Road, Newmarket, 1142 Auckland, New Zealand

**Abstract.** Polymer composites were manufactured using biocarbon particles as a reinforcing filler to improve the mechanical and thermal properties. However, a detailed examination of dispersion and agglomeration of filler is essential to correlate the filler/matrix and the filler/filler interactions with the mechanical properties of the product. We investigated the variations of mechanical, agglomeration behavior of fillers, and thermal properties of polypropylene (PP)/coconut shell biocarbon (CSB) composites. PP/CSB composites were prepared by melt mixing process varying the CSB content (0 to 20 wt%) using a Brabender mixer. The nanomechanical mapping of the composites studied using Atomic Force Microscopy revealed an increase in Young's modulus from 1.6 to 2.9 GPa when CSB loading increased from 0 to 20 wt%. The dispersion and agglomeration of CSB filler in the PP matrix were investigated using 3D reconstructed images with the help of X-ray micro-CT, and a dedicated 3D reconstruction software. The thermal stability of the PP/CSB composites also improved with an increase in CSB content in PP.

## 1. Introduction

Globally, carbon-based polymer composites have high demands owing to their advantages, such as low cost, lightweight, corrosion-resistance, low processing temperatures, and recyclability. A significant number of studies have been conducted using carbonaceous fillers (*e.g.*, carbon black [1–3], graphene [4–7], carbon nanotube [8, 9] and biocarbon [10, 11]) for enhancing the mechanical properties of the resultant composites. Although carbon-based fillers have numerous advantages, most of the commonly used carbon-based fillers are not essentially manufactured from renewable sources. For example, carbon black is mostly produced by burning petrochemicals in a controlled atmosphere [12].

While considering the potential impact of petroleum-based fillers such as carbon black and their composites on the environmental and climatic changes, it is rational to replace them with sustainable fillers. One of such sustainable and renewable carbon-based filler for the polymer composites is biocarbon, obtained from the biowaste [13]. Biocarbon is a porous carbonaceous material produced by the pyrolysis of biomass in a limited or oxygen-free atmosphere. Although biocarbon with different carbon contents and functionalities can be produced with selected manufacturing conditions, biocarbon with high carbon content and low concentration of functional groups are obtained by the pyrolysis of biomass at high temperatures in an inert atmosphere [14]. Pyrolysis of biomass at higher temperatures (>700 °C) improves the porosity as the volatile components of the biomass escape from the bulk, leaving behind a highly porous skeleton of carbonaceous material [14].

*Corresponding author, e-mail: jgeo541@aucklanduni.ac.nz
© BME-PT

In this study, the biocarbon used was produced from coconut shells. Coconut shell has been considered as a biowaste from the coconut tree (*Cocos nucifera*), traditionally used to produce activated carbon and finds many advanced applications in recent years, such as wastewater treatment [15], absorbent [16, 17], supercapacitors [18] and many more [19, 20]. Therefore, the coconut shell could be a suitable candidate for biocarbon preparation via pyrolysis in an inert atmosphere [21]. One of the significant advantages of using coconut shell biocarbon as a filler in polymer composites is the reduction of environmental pollution by maximizing the use of renewable sources. From the previous studies, it is understood that the mechanical properties of the composites can be tuned by carefully selecting the type of biocarbon and the filler loading concentration [14, 22].

Although several studies emphasize the utilization of naturally available bio-based reinforcing material, one of the major challenges is its incompatibility with the polymer matrix due to the presence of polar groups [23]. However, the compatibility of the bio-reinforced filler with the matrix can be improved by the addition of compatibilizers [24]. As the biocarbon produced at higher temperatures (>700 °C) is devoid of polar groups [14], the additional requirement for compatibilizers is not essential. The absence of polar groups on the filler surface can enhance a better interfacial interaction between the polymer matrix and the filler material. Therefore, the usage of such a filler can reduce the additional cost of a compatibilizer [25].

Some of the parameters that can influence the mechanical properties of the polymer composites are the filler geometry, the aspect ratio of a filler and its orientation in the polymer matrix, the surface chemistry of the filler material and its interaction with the polymer chains, the filler/filler interaction and the dispersion of the fillers. Based on these parameters and other processing conditions, the fillers can get well dispersed or agglomerated in a polymer matrix. One of the powerful characterization to understand the filler/filler and the filler/matrix interactions is rheology [26]. Other conventional methods to evaluate the filler/matrix interaction is electron microscopy (Transmission Electron Microscopy and Scanning Electron Microscopy). However, TEM reveals the 2D projection of a thin sample layer [27] while SEM provides the surface morphology [28], and it cannot truly represent the information in bulk. Therefore, a suitable technique called X-ray computer tomography is introduced to understand the bulk properties and filler dispersion in the polymer composites. X-ray micro-CT is a versatile technique to understand the bulk properties of the polymer composites [29]. The capabilities of X-ray computer tomography is utilized for the investigation and visualization of the dispersion of fillers in a bulk composite sample. Further, micro-CT is used to evaluate the filler dispersion and the agglomeration in the bulk matrix by 3D reconstruction of the 2D images of the bulk sample.

Besides, the viscoelastic properties of the samples can be extracted by AM-FM (amplitude modulation frequency modulation) mode present in atomic force microscopy (AFM). The advantage of using AFM in AM-FM mode is its capability in acquiring a viscoelastic map of the sample with minimal sample preparation. To the best of our knowledge, there has been no substantial study that investigates the nanomechanical behavior of PP/coconut shell biocarbon (CSB) composites using AFM (AM-FM) and Micro-CT techniques for the analysis of dispersion and agglomeration of CSB filler in PP matrix. In summary, the aim of the work is to maximize the utilization of coconut shell biocarbon produced at high temperatures to develop a novel biocomposite and investigate the nanomechanical behavior, dispersion, agglomeration of biocarbon in the polymer matrix and the thermal properties using advanced characterization techniques.

## 2. Materials and methods
### 2.1. Materials
The coconut shell biocarbon (CSB) was produced by pyrolysis of coconut shell at 800 °C in a nitrogen atmosphere (COCO GAC, supplied by Jacobi Carbon, China). The CSB supplied was dried, milled and sieved to under 100 μm and stored in a desiccator until it is used for the mixing. The average size of the particles was calculated using Mastersizer 2000 laser particle size analyzer (Malvern Panalytical,

U.K), and the SEM images were acquired to confirm the particle size. Polypropylene was supplied by LyodellBasell Mophlen HP 400 N with a melt flow rate (MFR) of 11 g/10 min (ISO 1183) and a density of 0.9 g/cm$^3$ supplied by TCL HUNT (New Zealand). The surface morphology and elemental analysis were studied using Hitachi SU-70 FE-SEM equipped with an energy-dispersive X-ray (EDX) facility, Japan. SEM analysis of polymer composites was used to evaluate the filler distribution and the interaction of the fillers with the polymer matrix. In order to avoid charging, the samples were coated with platinum, and a low acceleration voltage of 5 kV was used for imaging. The particle size distribution of CSB was conducted using Mastersize 2000 particle size analyzer. The Raman spectrum of biochar was carried out using Renishaw RM 1000 Raman Microprobe (UK) with an air-cooled argon-ion 488 nm excitation laser. XRD analysis was performed using a Rigaku Ultima IV diffractometer (Japan) with monochromatic Cu $K_{\alpha 1}$ radiation with wavelength 0.154059292 nm at 40 kV, 30 mA, and 1.2 kW. The diffraction data for the CSB sample was collected in the 2θ range (20–60°) at a scan rate of 2° 2θ/min.

## 2.2. CSB/PP composites preparation and characterisation

CSB/PP composites with different formulations of coconut shell biocarbon (5, 10, 15, and 20% w/w) were prepared by melt blending technique in Brabender Plastic-Corder Lab-station. The CSB and PP were dried in an oven for 6 hours and weighed and premixed according to the formulation mentioned earlier. The premixed CSB/PP was introduced into a 50 cm$^3$ melt-mixing chamber kept at 210 °C, and the mixing was done for 3 minutes with a screw rotation speed of 70 rpm.

After the specified mixing time, the CSB/PP composites were collected, pelletized by a pelletizing mill (Model: IZ-120, Lab tech Engineering Company Ltd, Thailand). The pellets were dried at 72 °C for 24 hours and pressed into flat sheets of (10 cm × 10 cm × 0.5 cm) size and used for the nanomechanical characterization. The pellets were also used to prepare dog bones for tensile experiments in an injection molding machine (Dr. Boy, GMBH 53577, Neustadt, Germany). The composites were also hot-pressed to a 6 mm diameter × 20 mm height cylindrical solid for the Micro-CT analysis. Differential scanning calorimetry (DSC) studies were performed using DSC Q 2000 (TA instrument, New Castle, DE, USA). The heating and cooling program was from 30 to 220 °C at a rate of 10 °C/min and a holding time of 4 minutes. To evaluate the interaction of CSB with PP, attenuated total reflectance Fourier transform infrared spectroscopy (ATR-FTIR) of the composites was carried out using Thermo Scientific Nicolet iS50 FT-IR with a diamond micro-tip accessory (400–4000 cm$^{-1}$). Thermogravimetric analysis (TGA) of the composites was performed in TGA Q 5000 (TA instrument, New Castle, DE, USA). The temperature program was set from 30 to 750 °C at a rate of 10 °C/min under nitrogen atmosphere (flow rate 50 ml/min).

Mechanical testing studies were performed on an Instron 5567 (USA) under ambient conditions in accordance with ASTM D 638 protocol, and a video extensometer was used for measuring the extension.

Micro-CT (Sky scan 1272, Bruker co, Billerica, MA, USA) was used to measure the dispersion and agglomeration of filler materials in the composites [30]. The non-destructive analysis was performed at a voltage of 48 kV and a current of 200 μA with a pixel size of 2452×1640 under normal pressure. The 2D images were sliced and reconstructed in three dimensions using a 3D reconstruction software (Amira). The dispersion of CSB and the volume of the agglomerates were calculated using Amira.

The nanomechanical studies of the composites were carried out using Environmental Atomic Force Microscope cypher ES (Asylum Research, Oxford Instruments). Amplitude Modulation-Frequency Modulation (AM-FM) viscoelastic mapping was used to characterize the viscoelastic properties of the resultant composites under bimodal tapping mode. The AFM cantilever, AC160TSA-R3 made of Silicon (Cr/Au reflex coating), with a tetrahedral silicon tip from Olympus was used. The spring constant and resonance frequency of the tip was 26 N/m and 300 kHz, respectively. The radius of the tip was 7 nm and a height of 14 μm. The AFM images and data were examined using IGOR pro® software.

## 3. Results and discussion
### 3.1. Coconut shell biocarbon (CSB) characterization

SEM images, Figure 1, revealed that the CSB particles were irregular in shape with particle size ranging from 1 to 40 μm. Figure 2 shows the Energy Dispersive Spectroscopy (EDS) spectrum of CSB, which shows that the major element present in the CSB is carbon with 95.92 wt% and the traces of other elements.

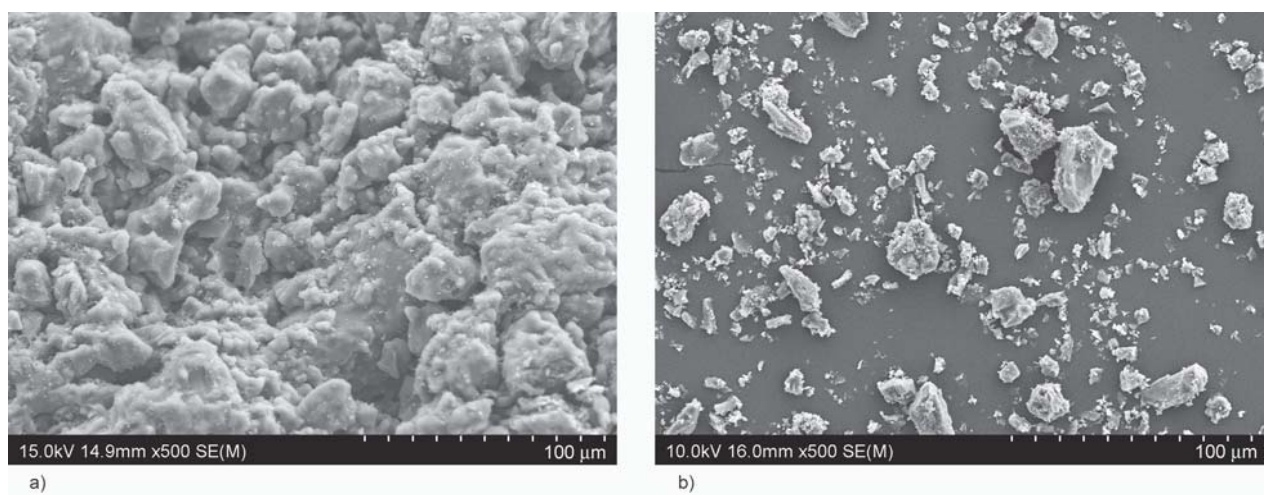

**Figure 1.** SEM images of CSB (a) before milling, and (b) after milling.

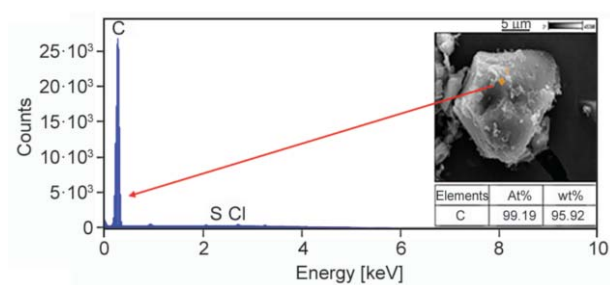

**Figure 2.** Energy dispersive spectroscopy (EDS) analysis of CSB.

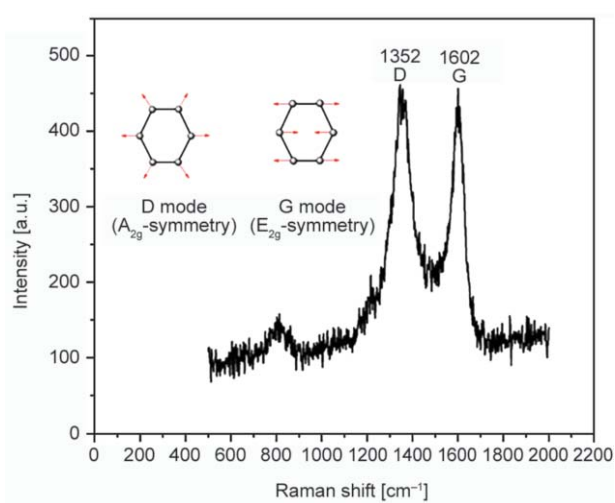

**Figure 3.** Raman spectrum of CSB.

The particle size distribution of the coconut shell biocarbon particles dispersed in water shows 10% of the particles are less than 4.575 μm particle diameter, 50% of the particles are less than 15.797 μm particle diameter, and 90% of the particles are less than 37.015 μm particle diameter. The specific surface area is about 0.6 m$^2$/g. The mean diameter of the CSB particles is 18.51 μm, and the diameter range is from 1.09 to 52.48 μm.

Figure 3 displays the Raman spectra of CSB. The main characteristics of the Raman spectra of carbon with hexagonal lattice are the G ($E_{2g}$ symmetry) and D ($A_{1g}$ symmetry) bands. The peak at 1602 cm$^{-1}$ correlated with G band, which is linked with the bond stretching of sp$^2$ hybridized carbon in the ring. In the case of rice husk biochar produced at 500 °C, G band appears at 1604 cm$^{-1}$ [31]. The peak at 1352 cm$^{-1}$ assigns to the D band, which is due to the breathing motion of aromatic rings [32, 33]. For graphite crystals, the G vibration mode is at 1589 cm$^{-1}$. The inset shows the representation of vibrational modes (G and D bands). According to the position of G band, the graphene layer stacking arrangement in CSB is low and defective. In the case of perfect graphitic crystal and graphene, the D band is absent but appears when the disordered structure is present. Due to the high pyrolysis temperature and milling process, defects can arise in CSB, and the D band is prominent in such defective carbon clusters [34, 35]. The $I_D/I_G$ ratio of the CSB is 1.55, which shows that the graphene layers in the CSB have a higher amount of defects.

The XRD pattern of CSB in Figure 4 shows a broad peak (002) at 2θ = 25.5°, which is attributed to the (0 0 2) reflection from the disordered carbon and a weak reflection at 43.3° corresponding to (1 0 0) the graphene peak and interlayer condensation. The sharp peaks of CSB at 2θ values of 26.5° (002), 42.3° (020), and 50.1° (022) correspond to the ordered graphitic structure [36]. The pyrolysis of coconut shells at high temperatures leads to the formation of ordered and disordered structures in CSB [37, 38], which may be due to the special processing conditions of the CSB.

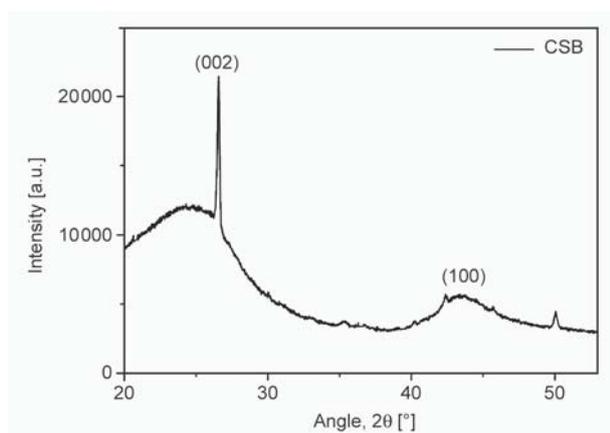

**Figure 4.** X-ray Diffraction analysis of CSB.

### 3.2. SEM analysis of PP/CSB composites

Figure 5 shows the SEM analysis of CSB/PP composites with different biocarbon loading concentrations. Figure 5a displays the neat PP without CSB filler and Figure 5b shows the PP with 5 wt% of CSB with no visible agglomeration of CSB. The agglomerations of CSB particles start with a higher concentration of the filler (PP-10CSB to PP-20CSB). Figure 5c reveals that the interaction of CSB with PP is weak, which is inferred from the gap in the interfacial region; note that it is a common phenomenon in the case of carbon-based composites [2, 39].

### 3.3. Mechanical characteristics of PP/CSB composites

The tensile properties of neat PP and PP/CSB composites are shown in Figure 6. The incorporation of biocarbon slightly decreases the tensile strength of the composites. The neat PP shows an average tensile strength of 35 MPa, and there is no considerable decrease in the tensile strength of composites with the addition of 10 wt% of CSB to the PP matrix. However, with the incorporation of 20 wt% CSB to the PP matrix, the tensile strength of the composites decreased to 33.7 MPa, which is 3.7% lower than that of neat polymer. The decrease in tensile strength was observed in other biocarbon based composites [14]. This decrease in the tensile strength of the composites can be attributed to the weak bhteraction of cSB particles with the PP matrix, which is evident from the SEM images (Figure 5) of PP/CSB composites.

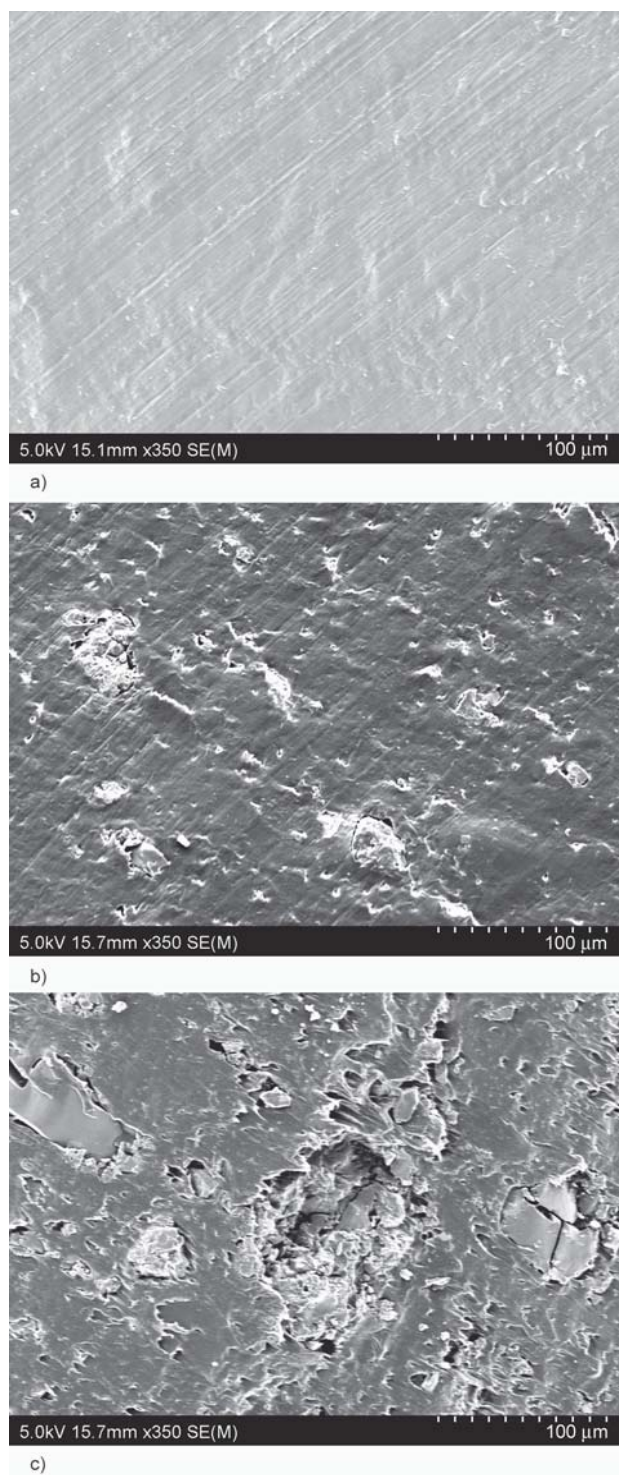

**Figure 5.** SEM images of (a) PP (b) PP-5CSB (c) PP-20CSB.

Figure 6 illustrates the elongation at yield of PP/CSB composites. The elongation at the yield of PP/CSB gradually decreases with the increasing concentration of CSB particles in the PP matrix. The substantial decrease in polymer ductility can be understood as the addition of CSB particles create networks in the polymer matrix, which reduces the mobility of the polymer chain in the composites [40].

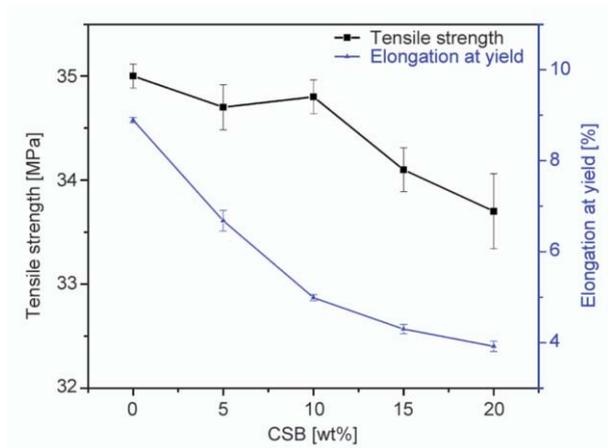

**Figure 6.** Tensile strength and elongation at yield of PP/CSB composites with varying biocarbon percentage.

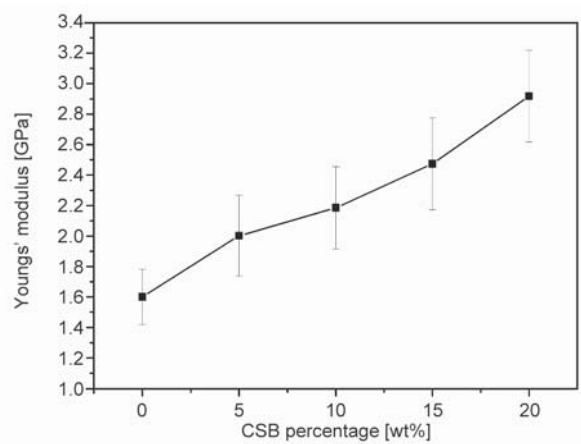

**Figure 8.** Variation of Young's modulus of the PP composites with the weight percentage of CSB.

### 3.4. Nanomechanical mapping of PP/CSB composites

In this study, two different frequencies were used to obtain a topography and viscoelastic properties of CSB/PP composites. In AM-FM mode, the topographical feedback is carried out by amplitude mode, which is confined to the first resonance frequency, and second excitation mode extracts Young's moduli of the composites by frequency shifts compared to the resonance frequency. At first, a reference sample with specific Young's modulus was used to calibrate the cantilever response. The effective Young's modulus of the sample can be derived as, $E_{eff} = C_2 \cdot \Delta f_2$, where $C_2$ is measured over the reference sample; this can be applied to unknown samples for necessary calculations. A detailed explanation of the AM FM mode is provided in reference [41].

Figure 7 illustrates the distribution of Young's modulus values for PP and PP-20CSB. The red line displays the distribution (Gaussian) of Young's modulus values of PP, which is narrow, indicating the homogeneity of the PP film. The blue line indicates the distribution of Young's modulus values of PP-20CSB. The spread in Young's modulus values in the case of PP-20CSB is higher compared to that for PP, and the broadening of the distribution indicates the inhomogeneity of the composites.

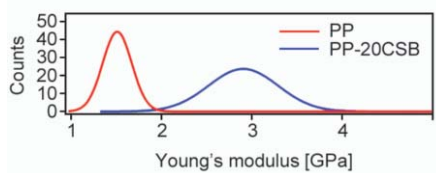

**Figure 7.** Young's modulus of PP and PP-20CSB using AFM.

Figure 8 presents the variation of Young's modulus of the polymer composites with the increase in weight percentage of CSB in the PP matrix. The Young's moduli of the composites increase with an increase in CSB concentration from 0 to 20 wt%, which is observed in many cases of carbon-based polymer composites [14, 42]. The increase in Young's moduli of composites implies that the introduction of carbon-based fillers in the polymer matrix reduces the polymer chain movement, which increases the resistance of the deformation of the composites [40].

### 3.5. X-ray micro-CT analyses

X-ray micro-CT is a non-destructive technique to understand the internal geometry and distribution of particles in composite materials [43]. Although it is a popular technique in medical imaging, the use of micro-CT for biocarbon composites has not been explored yet. The micro-CT scan gives good contrast images if the X-ray attenuation of the filler materials is higher compared to the matrix. Here the micro-CT images gave a better contrast due to the density difference between the CSB filler and PP matrix. Figure 9 shows the micro-CT image of the distribution of CSB filler in a 3D cubic volume of $1000 \times 1000 \times 1000\ \mu m^3$. Figure 9a illustrates the 3D reconstructed image of the PP-5CSB, which reveals that the CSB particles are homogeneously distributed in PP with low concentrations of agglomerations. Figure 9b–9d show the 3D reconstructed images of PP-10CSB, PP-15CSB and PP-20CSB, respectively. When the concentration of the CSB filler material increases, the agglomeration of the particles increases, which attributes to the non-covalent interactions between the filler particles [44, 45] and becomes dominant when the surface of contact of fillers with polymer matrix is reduced. The interfacial adhesion of filler with polymer matrix plays a major role in the agglomeration of fillers in the composites. The agglomeration of fillers can lead to the formation of networks (rheological percolation threshold) in the composite systems, which can enhance the storage modulus of the system by hindering the free movement of the polymer chains [2, 40].

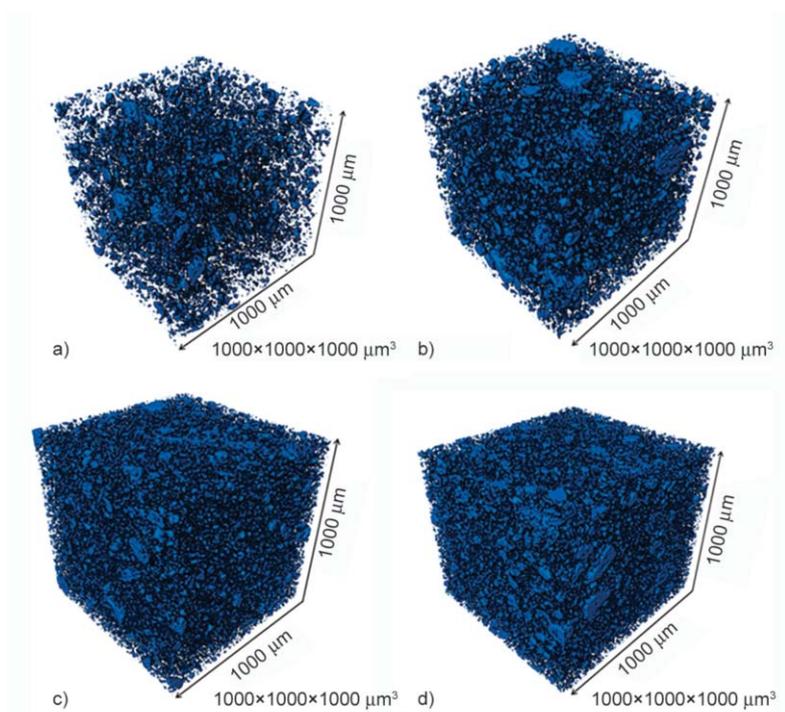

**Figure 9.** Micro-CT CSB particles distribution of (a) PP-5CSB (b) PP-10CSB (c) PP-15CSB (d) PP-20CSB.

Figure 10 displays the number of CSB particles vs. filler material volume in micrometer, and as expected, the number of particles with larger volume (agglomerated particles) increases with the increase in filler loading concentration. In order to investigate the amount and size of the agglomeration, the volume of the largest single particle is calculated from the particle size analysis data. The volume of the largest single particle is 75 300 μm$^3$ (assuming particle as spherical), and any particle volume above this can be considered as agglomeration. The results show that the agglomeration is low in the case of PP-5CSB and maximum in the case of PP-20CSB, (Figure 10). The volume of agglomerates also increases with an increase in the loading concentration of the filler. The maximum volume of the agglomerate in the case of PP-5CSB is 98 423 μm3, which increases to 12 524 400 μm$^3$ in the case of PP-20CSB. Thus, the data from the micro-CT would be useful for the understanding of the agglomeration behavior of CSB and other micron-sized particles in a system.

### 3.6. ATR-FTIR analysis of PP/CSB composites

The ATR-FTIR, Figure 11, experiments were conducted to investigate the presence of any functional groups in CSB and the chemical interaction of the biocarbon with the polymer matrix. The FTIR spectrum of CSB revealed the absence of any functional groups on the surface of biocarbon particles. In the FTIR spectrum of PP/CSB composites, the absence of new peaks is observed while existing peaks remain the same, which confirms that the mixing is purely a physical process. The sharp absorption peaks observed at 2972 cm$^{-1}$ can correspond to –CH$_3$ asymmetric stretching vibrations, peaks at 985, and 1165 cm$^{-1}$ are assigned to –CH$_3$ rocking vibrations and the absorption peak at 1360 cm$^{-1}$ corresponds to symmetric bending vibration of –CH$_3$ group. The peaks at 1450, 2840 and 2920 cm$^{-1}$ correspond to –CH$_2$ symmetric bending, –CH$_2$ symmetric stretching and -CH$_2$ asymmetric stretching, respectively [14, 46].

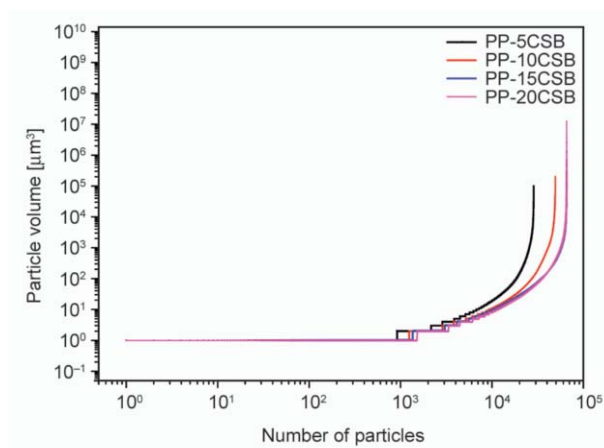

**Figure 10.** Biocarbon particle volume distribution.

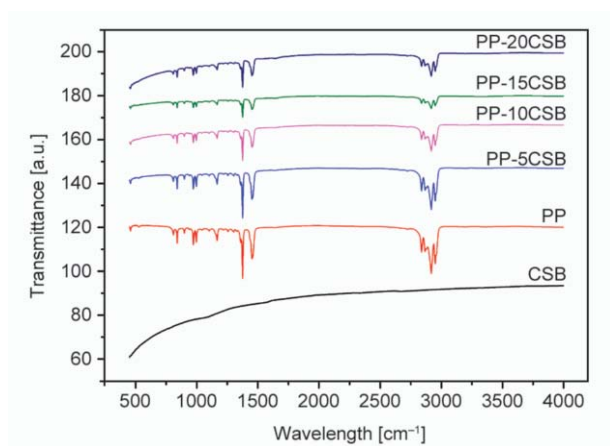

**Figure 11.** FTIR of coconut shell and coconut shell biocarbon (CSB).

**Table 1.** DSC data on PP/CSB composites.

| Material | $T_c$ [°C] | $T_m$ [°C] | $\Delta H_m$ [J/g] | $X_c$ [%] |
|---|---|---|---|---|
| PP | 120.95 | 152.37 | 99.71 | 48.17 |
| PP-5CSB | 125.66 | 152.69 | 92.64 | 47.11 |
| PP-10CSB | 126.37 | 154.82 | 85.36 | 45.63 |
| PP-15CSB | 127.62 | 152.31 | 78.98 | 44.54 |
| PP-20CSB | 130.03 | 154.44 | 64.36 | 38.86 |

### 3.7. Thermal behavior of PP/CSB composites

The thermograph of PP, Figure 12, exhibits a melting peak at 152.37 °C and a melt crystallization peak at 120.95 °C. The presence of CSB in the PP matrix shifts the crystallization temperature to a higher temperature compared to that of neat PP, Figure 12, which is due to the nucleating effect of CSB particles in the PP matrix where biocarbon particles act as nucleating sites, the phenomenon that is explained elsewhere [42]. Upon increasing the biocarbon content from 0 to 20 wt% the $\Delta H_m$ values decrease from 99.71 to 64.36 J/g, respectively. The degree of crystallinity ($X_c$) of the polymer composites can be calculated by using Equation (1):

$$X_c [\%] = \frac{\Delta H_m}{(1 - \varphi)\Delta H_m^0} \cdot 100 \quad (1)$$

where ($\varphi$) is the weight fraction of CSB in the composites, ($\Delta H_m$) is the enthalpy of melting, and ($\Delta H_m^0$) is the enthalpy of melting of 100% crystalline PP which is found to be 207 J/g [47]. The addition of biocarbon reduces the total crystallinity of the composites, Table 1. The total crystallinity of PP decreases from 48.17 to 38.86% when CB loading was increased from 5 to 20 wt% of CSB loading. The rate of decrease in crystallinity is significant when the loading of CSB filler is higher than 15 wt%. The crystallinity of PP decreases sharply when the loading of CSB increases from 15 to 20 wt%, which is due to the restricted motion of PP chains in the presence of CSB. The mobility of polymer chains within and around the aggregates are restricted due to network formation of filler particles in the polymer matrix. This network formation makes the polymer chain rearrangement difficult and leads to a decrease in the degree of crystallinity with the increase in CSB loading concentration [48, 49]. The CSB particles can occupy the interstitial positions of the PP chains, which reduces the orderly arrangements and free volume for the motion of the polymer chains; hence the crystallinity is reduced with an increase in filler concentration [14].

Figure 13 illustrates the TGA, Figure 13a, and DTG, Figure 13b, of PP and PP/CSB. The onset of thermal degradation of the neat PP occurs at 428 °C, but with the incorporation of CSB as a filler, the onset of degradation shifts to a higher temperature (480 °C), which is about 12% higher than that of neat PP sample. There is an abrupt improvement in the thermal stability of the composites with 5 wt% loading of CSB to the PP matrix, but the relative increase in the onset of degradation is lower with higher loading of CSB. In general, the addition of CSB to the PP improves the thermal stability of the composites. As the biocarbon is produced at higher temperatures with a high percentage of graphitic carbon, it is thermally stable compared to the PP matrix. While comparing the residue at the end of the curve, a composite with a higher percentage of CSB (PP-20CSB) shows a larger amount of residue [31]. The improvement in the thermal stability of composites can be due to the high thermal conductivity and efficient dissipation of thermal energy by the biocarbon particles to the surrounding matrix [50].

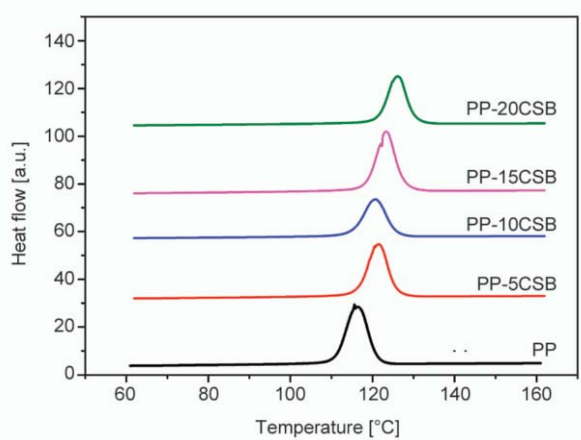

**Figure 12.** Crystallisation curve of PP/CSB composites.

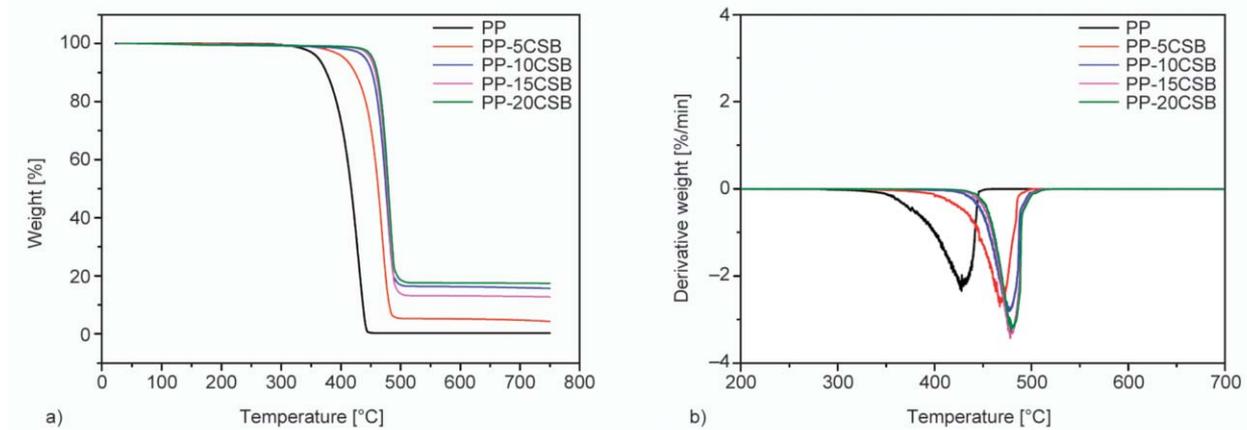

**Figure 13.** (a) TGA and (b) DTG curve of PP/CSB composites.

## 4. Conclusions

Through this study, nanomechanical properties (Young's modulus mapping) of PP/ CSB composites have been investigated using dynamic mode in AFM. Young's modulus of a composite increased from 1.6 GPa (PP) to 2.9 GPa (PP-20CSB) with an increase in CSB concentration in PP. The introduction of the X-ray micro-CT technique has added a novel way of understanding the dispersion and agglomeration of CSB in PP. Along with SEM, 3D reconstructed micro-CT images of the filler particles in the polymer matrix can give a new insight into the particle distribution and agglomeration behavior of CSB in PP. 3D reconstructed micro-CT images reveal that the CSB-CSB interaction, which leads to the agglomeration of biocarbon, is the dominant factor compared to CSB-PP interaction. The DSC results reveal that the CSB acts as nucleating sites in PP for crystallization. The TGA results indicate that the thermal stability of the resultant composites can improve with the addition of CSB. Further work is being carried out to improve the dispersion of biocarbon in a polymer matrix, which can enhance the filler-matrix interaction and reduce the agglomerations.

**Acknowledgements**
The authors thank all technicians at the Centre for Advanced Composites Materials (CACM) for their assistance. In addition, the authors would like to acknowledge the Ministry of Business, Innovation & Employment, New Zealand, for their financial support (grant number: UOA 3706657).

**References**

[1] Burmistrov I., Gorshkov N., Ilinykh I., Muratov D., Kolesnikov E., Anshin S., Mazov I., Issi J-P., Kusnezov D.: Improvement of carbon black based polymer composite electrical conductivity with additions of MWCNT. Composites Science and Technology, **129**, 79–85 (2016).
https://doi.org/10.1016/j.compscitech.2016.03.032

[2] Poulose A. M., Anis A., Shaikh H., George J., Al-Zahrani S. M.: Effect of plasticizer on the electrical, thermal, and morphological properties of carbon black filled poly(propylene). Polymer Composites, **38**, 2472–2479 (2017).
https://doi.org/10.1002/pc.23834

[3] Zhao Y-H., Zhang Y-F., Wu Z-K., Bai S-L.: Synergic enhancement of thermal properties of polymer composites by graphene foam and carbon black. Composites Part B: Engineering, **84**, 52–58 (2016).
https://doi.org/10.1016/j.compositesb.2015.08.074

[4] Das T. K., Prusty S.: Graphene-based polymer composites and their applications. Polymer-Plastics Technology and Engineering, **52**, 319–331 (2013).
https://doi.org/10.1080/03602559.2012.751410

[5] Kuilla T., Bhadra S., Yao D., Kim N. H., Bose S., Lee J. H.: Recent advances in graphene based polymer composites. Progress in Polymer Science, **35**, 1350–1375 (2010).
https://doi.org/10.1016/j.progpolymsci.2010.07.005

[6] Potts J. R., Dreyer D. R., Bielawski C. W., Ruoff R. S.: Graphene-based polymer nanocomposites. Polymer, **52**, 5–25 (2011).
https://doi.org/10.1016/j.polymer.2010.11.042

[7] Li Y., Feng Z., Huang L., Essa K., Bilotti E., Zhang H., Peijs T., Hao L.: Additive manufacturing high performance graphene-based composites: A review. Composites Part A: Applied Science and Manufacturing, **124**, 105483/1–105483/22 (2019).
https://doi.org/10.1016/j.compositesa.2019.105483

[8] Valentini L., Kenny J. M.: Novel approaches to developing carbon nanotube based polymer composites: Fundamental studies and nanotech applications. Polymer, **46**, 6715–6718 (2005).
https://doi.org/10.1016/j.polymer.2005.05.025

[9] Castellino M., Rovere M., Shahzad M. I., Tagliaferro A.: Conductivity in carbon nanotube polymer composites: A comparison between model and experiment. Composites Part A: Applied Science and Manufacturing, **87**, 237–242 (2016).
https://doi.org/10.1016/j.compositesa.2016.05.002

[10] Nan N., DeVallance D. B., Xie X., Wang J.: The effect of bio-carbon addition on the electrical, mechanical, and thermal properties of polyvinyl alcohol/biochar composites. Journal of Composite Materials, **50**, 1161–1168 (2016).
https://doi.org/10.1177/0021998315589770

[11] Zhang Q., Khan M. U., Lin X., Cai H., Lei H.: Temperature varied biochar as a reinforcing filler for high-density polyethylene composites. Composites Part B: Engineering, **175**, 107151/1–107151/7 (2019).
https://doi.org/10.1016/j.compositesb.2019.107151

[12] Fulcheri L., Probst N., Flamant G., Fabry F., Grivei E., Bourrat X.: Plasma processing: A step towards the production of new grades of carbon black. Carbon, **40**, 169–176 (2002).
https://doi.org/10.1016/S0008-6223(01)00169-5

[13] Antal M. J., Wade S. R., Nunoura T.: Biocarbon production from Hungarian sunflower shells. Journal of Analytical and Applied Pyrolysis, **79**, 86–90 (2007).
https://doi.org/10.1016/j.jaap.2006.09.005

[14] Poulose A. M., Elnour A. Y., Anis A., Shaikh H., Al-Zahrani S., George J., Al-Wabel M. I., Usman A. R., Ok Y. S., Tsang D. C., Sarmah A. K.: Date palm biochar-polymer composites: An investigation of electrical, mechanical, thermal and rheological characteristics. Science of The Total Environment, **619–620**, 311–318 (2018).
https://doi.org/10.1016/j.scitotenv.2017.11.076

[15] Amuda O. S., Giwa A. A., Bello I. A.: Removal of heavy metal from industrial wastewater using modified activated coconut shell carbon. Biochemical Engineering Journal, **36**, 174–181 (2007).
https://doi.org/10.1016/j.bej.2007.02.013

[16] Freitas J. V., Nogueira F. G. E., Farinas C. S.: Coconut shell activated carbon as an alternative adsorbent of inhibitors from lignocellulosic biomass pretreatment. Industrial Crops and Products, **137**, 16–23 (2019).
https://doi.org/10.1016/j.indcrop.2019.05.018

[17] Abe I., Fukuhara T., Maruyama J., Tatsumoto H., Iwasaki S.: Preparation of carbonaceous adsorbents for removal of chloroform from drinking water. Carbon, **39**, 1069–1073 (2001).
https://doi.org/10.1016/S0008-6223(00)00230-X

[18] Mi J., Wang X-R., Fan R-J., Qu W-H., Li W-C.: Coconut-shell-based porous carbons with a tunable micro/mesopore ratio for high-performance supercapacitors. Energy and Fuels, **26**, 5321–5329 (2012).
https://doi.org/10.1021/ef3009234

[19] Noori A., Bartoli M., Frache A., Piatti E., Giorcelli M., Tagliaferro A.: Development of pressure-responsive polypropylene and biochar-based materials. Micromachines, **11**, 339/1–339/12 (2020).
https://doi.org/10.3390/mi11040339

[20] Bartoli M., Giorcelli M., Jagdale P., Rovere M., Tagliaferro A. J. M.: A review of non-soil biochar applications. Materials, **13**, 261/1–261/35 (2020).
https://doi.org/10.3390/ma13020261

[21] Jain A., Aravindan V., Jayaraman S., Kumar S. P., Balasubramanian R., Ramakrishna S., Madhavi S., Srinivasan M. P.: Activated carbons derived from coconut shells as high energy density cathode material for Li-ion capacitors. Scientific Reports, **3**, 3002/1–3002/6 (2013).
https://doi.org/10.1038/srep03002

[22] Das O., Sarmah A. K., Bhattacharyya D.: Biocomposites from waste derived biochars: Mechanical, thermal, chemical, and morphological properties. Waste Management, **49**, 560–570 (2016).
https://doi.org/10.1016/j.wasman.2015.12.007

[23] Li X., Tabil L. G., Panigrahi S.: Chemical treatments of natural fiber for use in natural fiber-reinforced composites: A review. Journal of Polymers and the Environment, **15**, 25–33 (2007).
https://doi.org/10.1007/s10924-006-0042-3

[24] Das O., Sarmah A. K., Bhattacharyya D.: A novel approach in organic waste utilization through biochar addition in wood/polypropylene composites. Waste Management, **38**, 132–140 (2015).
https://doi.org/10.1016/j.wasman.2015.01.015

[25] Behazin E., Misra M., Mohanty A. K.: Compatibilization of toughened polypropylene/biocarbon biocomposites: A full factorial design optimization of mechanical properties. Polymer Testing, **61**, 364–372 (2017).
https://doi.org/10.1016/j.polymertesting.2017.05.031

[26] Huang H-X., Zhang J-J.: Effects of filler–filler and polymer–filler interactions on rheological and mechanical properties of HDPE–wood composites. Journal of Applied Polymer Science, **111**, 2806–2812 (2009).
https://doi.org/10.1002/app.29336

[27] Qian D., Dickey E. C.: *In-situ* transmission electron microscopy studies of polymer–carbon nanotube composite deformation. Journal of Microscopy, **204**, 39–45 (2001).
https://doi.org/10.1046/j.1365-2818.2001.00940.x

[28] Kovacs J. Z., Andresen K., Pauls J. R., Garcia C. P., Schossig M., Schulte K., Bauhofer W.: Analyzing the quality of carbon nanotube dispersions in polymers using scanning electron microscopy. Carbon, **45**, 1279–1288 (2007).
https://doi.org/10.1016/j.carbon.2007.01.012

[29] Garcea S. C., Wang Y., Withers P. J.: X-ray computed tomography of polymer composites. Composites Science and Technology, **156**, 305–319 (2018).
https://doi.org/10.1016/j.compscitech.2017.10.023

[30] Kim H. S., Kim J. H., Yang C-M., Kim S. Y.: Synergistic enhancement of thermal conductivity in composites filled with expanded graphite and multi-walled carbon nanotube fillers *via* melt-compounding based on polymerizable low-viscosity oligomer matrix. Journal of Alloys and Compounds, **690**, 274–280 (2017).
https://doi.org/10.1016/j.jallcom.2016.08.141

[31] George J., Azad L. B., Poulose A. M., An Y., Sarmah A. K.: Nano-mechanical behaviour of biochar-starch polymer composite: Investigation through advanced dynamic atomic force microscopy. Composites Part A: Applied Science and Manufacturing, **124**, 105486/1–105486/9 (2019).
https://doi.org/10.1016/j.compositesa.2019.105486

[32] Ferrari A. C.: Raman spectroscopy of graphene and graphite: Disorder, electron–phonon coupling, doping and nonadiabatic effects. Solid State Communications, **143**, 47–57 (2007).
https://doi.org/10.1016/j.ssc.2007.03.052

[33] Ferrari A. C., Robertson J.: Interpretation of Raman spectra of disordered and amorphous carbon. Physical review B, **61**, 14095–14107 (2000).
https://doi.org/10.1103/PhysRevB.61.14095

[34] Tommasini M., Castiglioni C., Zerbi G., Barbon A., Brustolon M.: A joint Raman and EPR spectroscopic study on ball-milled nanographites. Chemical Physics Letters, **516**, 220–224 (2011).
https://doi.org/10.1016/j.cplett.2011.09.094

[35] Munir K. S., Qian M., Li Y., Oldfield D. T., Kingshott P., Zhu D. M., Wen C. J.: Quantitative analyses of MWCNT-Ti powder mixtures using Raman spectroscopy: The influence of milling parameters on nanostructural evolution. Advanced Engineering Materials, **17**, 1660–1669 (2015).
https://doi.org/10.1002/adem.201500142

[36] Zhao C., Wang Q., Lu Y., Li B., Chen L., Hu Y-S.: High-temperature treatment induced carbon anode with ultrahigh Na storage capacity at low-voltage plateau. Science Bulletin, **63**, 1125–1129 (2018).
https://doi.org/10.1016/j.scib.2018.07.018

[37] Zhao Q., Fellinger T-P., Antonietti M., Yuan J.: A novel polymeric precursor for micro/mesoporous nitrogen-doped carbons. Journal of Materials Chemistry A, **1**, 5113–5120 (2013).
https://doi.org/10.1039/C3TA10291B

[38] Liu X-Y., Huang M., Ma H-L., Zhang Z-Q., Gao J-M., Zhu Y-L., Han X-J., Guo X-Y. J. M.: Preparation of a carbon-based solid acid catalyst by sulfonating activated carbon in a chemical reduction process. Molecules, **15**, 7188–7196 (2010).
https://doi.org/10.3390/molecules15107188

[39] Kasaliwal G. R., Pegel S., Göldel A., Pötschke P., Heinrich G.: Analysis of agglomerate dispersion mechanisms of multiwalled carbon nanotubes during melt mixing in polycarbonate. Polymer, **51**, 2708–2720 (2010).
https://doi.org/10.1016/j.polymer.2010.02.048

[40] Poulose A. M., Elnour A. Y., Samad U. A., Alam M. A., George J., Sarmah A. K., Al-Zahrani S. M.: Nano-indentation as a tool for evaluating the rheological threshold in polymer composites. Polymer Testing, **80**, 106150/1–106150/8 (2019).
https://doi.org/10.1016/j.polymertesting.2019.106150

[41] Garcia R., Proksch R.: Nanomechanical mapping of soft matter by bimodal force microscopy. European Polymer Journal, **49**, 1897–1906 (2013).
https://doi.org/10.1016/j.eurpolymj.2013.03.037

[42] Das O., Bhattacharyya D., Hui D., Lau K-T.: Mechanical and flammability characterisations of biochar/polypropylene biocomposites. Composites Part B: Engineering, **106**, 120–128 (2016).
https://doi.org/10.1016/j.compositesb.2016.09.020

[43] Shen H., Nutt S., Hull D.: Direct observation and measurement of fiber architecture in short fiber-polymer composite foam through micro-CT imaging. Composites Science and Technology, **64**, 2113–2120 (2004).
https://doi.org/10.1016/j.compscitech.2004.03.003

[44] Jiang L. Y., Huang Y., Jiang H., Ravichandran G., Gao H., Hwang K. C., Liu B.: A cohesive law for carbon nanotube/polymer interfaces based on the van der Waals force. Journal of the Mechanics and Physics of Solids, **54**, 2436–2452 (2006).
https://doi.org/10.1016/j.jmps.2006.04.009

[45] Liao K., Li S.: Interfacial characteristics of a carbon nanotube–polystyrene composite system. Applied Physics Letters, **79**, 4225–4227 (2001).
https://doi.org/10.1063/1.1428116

[46] Gopanna A., Mandapati R. N., Thomas S. P., Rajan K., Chavali M.: Fourier transform infrared spectroscopy (FTIR), Raman spectroscopy and wide-angle X-ray scattering (WAXS) of polypropylene (PP)/cyclic olefin copolymer (COC) blends for qualitative and quantitative analysis. Polymer Bulletin, **76**, 4259–4274 (2019).
https://doi.org/10.1007/s00289-018-2599-0

[47] Paukkeri R., Lehtinen A.: Thermal behaviour of polypropylene fractions: 1. Influence of tacticity and molecular weight on crystallization and melting behaviour. Polymer, **34**, 4075–4082 (1993).
https://doi.org/10.1016/0032-3861(93)90669-2

[48] Nadiv R., Fernandes R. M., Ochbaum G., Dai J., Buzaglo M., Varenik M., Biton R., Furó I., Regev O. J. P.: Polymer nanocomposites: Insights on rheology, percolation and molecular mobility. Polymer, **153**, 52–60 (2018).
https://doi.org/10.1016/j.polymer.2018.07.079

[49] Seven K. M., Cogen J. M., Gilchrist J. F.: Nucleating agents for high-density polyethylene – A review. Polymer Engineering and Science, **56**, 541–554 (2016).
https://doi.org/10.1002/pen.24278

[50] Zhou W-Y., Qi S-H., Zhao H-Z., Liu N-L.: Thermally conductive silicone rubber reinforced with boron nitride particle. Polymer Composites, **28**, 23–28 (2007).
https://doi.org/10.1002/pc.20296